\documentclass{article}

\usepackage[a4paper]{geometry}

\usepackage[dvips]{graphicx}
\usepackage{color}

\usepackage{amssymb,amsfonts,amsmath}

\usepackage[dvips]{graphicx}
\usepackage{color}

\usepackage{amssymb,amsfonts,amsmath}

\def\degree{{k}}
\def\linksRC{k^+}

\def\weight{{w}}

\def\profile{{\alpha}}
\def\strength{{w}}

\def\nodeWeight{\sigma}
\def\nodeWeightPlus{\sigma^+}
\def\degree{k}

\def\nodeWeightPlusIn{\sigma^{+in}}
\def\nodeWeightPlusOut{\sigma^{+out}}

\begin{document}


\title{Rich-cores in networks}
\author{Athen Ma and Ra\'ul J Mondrag\'on\\
Queen Mary University of London, \\School of Electronic Engineering and Computer Science,\\ London, United Kingdom
}


\maketitle

\begin{abstract} 
A core is said to be a group of central and densely connected nodes which governs the overall behavior of a network. Profiling this meso--scale structure currently relies on a limited number of  methods which are often complex, and have scalability issues when dealing with very large networks. As a result, we are yet to fully understand its impact on network properties and dynamics. Here we introduce a simple method to profile this structure by combining the concepts of core/periphery and rich-club. The key challenge in addressing such association of the two concepts is to establish a way to define the membership of the core. The notion of a ``rich-club'' describes nodes which are essentially the hub of a network, as they play a dominating role in structural and functional properties. Interestingly, the definition of a rich-club naturally emphasizes high degree nodes and divides a network into two subgroups. Our approach theoretically couples the underlying principle of a rich-club with the escape time of a random walker, and a rich-core is defined by examining changes in the associated persistence probability.  The method is fast and scalable to large networks. In particular, we successfully show that the evolution of the core in \emph{C.~elegans} and World Trade networks correspond to key development stages and responses to  historical events respectively. 

\end{abstract}





 
%
%
Many networks exhibit a core/periphery dichotomy which is important for the understanding of network properties and dynamics (for a review see ref.~\cite{Csermely2013}). For instance, this meso--scale structure helps explain elitism in economic growth among countries \cite{krugman1995globalization}, organization of  the  human brain dynamics~\cite{bassett2012core} and inter--hub configurations in communication networks \cite{carmi2007model}. The constitution of a core/periphery structure often refers to a simple two-class partition \cite{Borgatti99,holme2005core, da2008centrality}. Members of the core are high degree nodes which have high inter-core and intra-partition linkages. In the original definition, the core/periphery was obtained by applying block-modeling to the adjacency matrix, where the connectivity within the periphery was assumed to be non-existent~\cite{Borgatti99}; however, this assumption is  too stringent for most real networks, and newer methods tend to relax this constraint. Other methods include using closeness centrality to minimize the diameter between the core and the rest of the network \cite{holme2005core, da2008centrality}, and applying Markov chains to describe random walks and provide an implicit partition \cite{della2013profiling}. Most studies focus on a single core, while the possibility of having a multi--core has been discussed \cite{Borgatti99, puck2012core}, the number of available methods for their profiling is rather limited. 

The notion of a rich-club is used to describe the connectivity between high degree nodes, and it has been applied to profile meso--scale properties in networks by examining the density of connections between high degree nodes \cite{Zhou04,Colizza06, Xu10, Heuvel11, towlson2013rich,mondragon2012}.  A rich-club influences the functionality of a network, as demonstrated in the transmission of rumors in social networks \cite{masuda2006vip}, the delivery of information in the Internet \cite{Zhou04} and its strong influence on both the network assortativity and transitivity \cite{Xu10}. Interestingly, the presence of a rich-club naturally divides a network into two parts.  This means that the rich-club concept coincides with two fundamental prerequisites of a core/periphery structure: \emph{high degree nodes} and \emph{two-class partition}. However, there is, at present, no general method to define the membership of the club~\cite{Xu10,towlson2013rich,valverde2007self}. 

Here, we present a \emph{rich--core} method which detects the core/periphery structure in complex networks, and formally defines the members of the associated rich-club. The method marks the formation of a rich-core by examining the persistence probability of random walks among high degree nodes. We analyzed a wide range of networks and successfully identified the core among them; our method also reveals multi-core within a core structure where one is present. Networks with a temporal nature have also been studied and our results show that the evolution of a core closely reflects the development and re--alignment of relationships over time in real networks. For example, changes in the core coincide with key development stages in a biological process and with timing of major historical events in trade development. In addition, anomalous nodes in networks can be uncovered with reference to null models by discriminating statistical differences. The rich-core method is simple, fast and applicable to very large networks. 

\section{Results and Discussion}
\begin{figure}[tb]
\begin{center}
\includegraphics{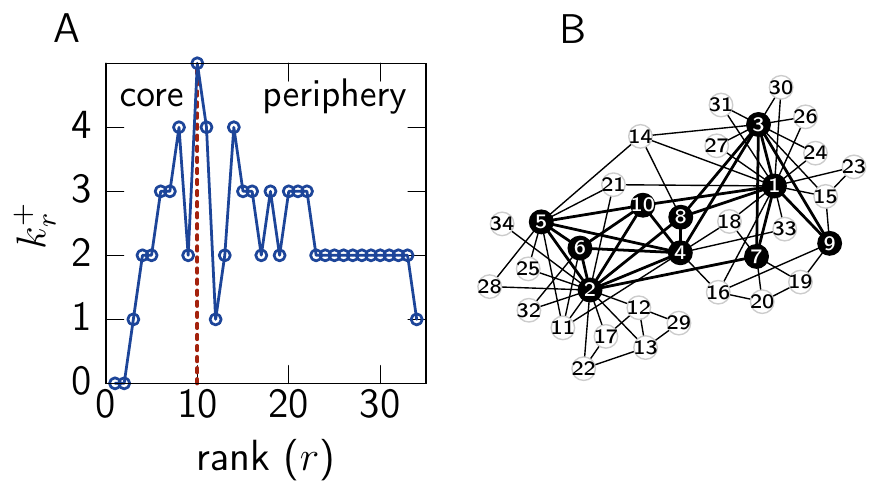}
\end{center}
\caption{\label{fig:defineCore2} The rich-core for the Zachary Karate Club network. (A) The number of links $\linksRC_r$ that node $r$ shares with nodes of a higher rank. The boundary of the core is marked by $\linksRC_{r^*}$ where $\linksRC_r$ is at its maximum, as indicated by the dotted line. (B) A graphical representation of the network with core (black) and periphery (white) nodes derived from the rich-core method.}
\end{figure}

Consider an unweighted and undirected graph. We rank the order of importance of nodes in descending order of their degree, such that the node with the highest degree is ranked first and so on. For a given node, we divide its links into two groups: those with nodes of a higher rank and those with a lower rank. More formally, a node with a rank $r$ has degree $\degree_r$; the number of links it shares with nodes of a higher rank is $\linksRC_r$ and the number of links with nodes of a lower rank is $\degree_r-\linksRC_r$. Core nodes are often high degree nodes that are densely connected with each other~\cite{Borgatti99}, and we assume that the connectivity of a highly ranked node with other \emph{higher} ranked nodes contributes towards the constitution of a core. Similarly, if such a node has very few links with higher ranked nodes, it is likely to be a member of the periphery.
To detect the core we propose this straightforward procedure (see Materials and Methods for full technical description). Starting from the node with the highest rank, as $r$ increases the number of links $\linksRC_r$ that node $r$ shares with nodes of a higher rank fluctuates. There will be a node $r^*$ where $\linksRC_r$ has reached its maximum, and from that node onwards $\linksRC_r$ is always less than $\linksRC_{r^*}$. This change in the connectivity among the highly ranked nodes is defined as the boundary of the (rich) core; the nodes with a rank less than or equal to $r^*$ are the core and the rest belong to the periphery. We studied the Zachary Karate Club network \cite{zachary1977information} and the rich-core is formed by the 10 highest ranked nodes (Fig.~\ref{fig:defineCore2}).  

This pragmatic way of defining a core is related to the concept of a random walker in a network. 
Consider a network which is partitioned into two sets: $S_c$ is the core and $S_p$ is the periphery. A random walker jumps from one node to another following a link between any pair of nodes, and the probability for the walker to visit a given node is proportional to the node's degree. The time it takes a random walker to escape from $S_c$ to $S_p$ is $\tau_c$. If $S_c$ is the set of nodes with the highest rank, as we include another node into this set the escaping time will naturally increase. Eventually, $S_c$ will contain all the nodes in the network, and the escaping time will diverge as the random walker is always contained in this set and has nowhere to escape to. This means that if we begin by putting only the top ranked node in set $S_c$, and gradually increase the size of $S_c$ by adding nodes in decreasing order of their rank, the escaping time will always increase.  If we consider the rate of change
of the escaping time as nodes are being added to the set, the boundary of a rich-core depicts the point in which the rate of increase on the escaping time changes from a fast to a slow pace (Materials and Methods). 

This definition of a rich-core can be easily extended to weighted networks~\cite{serrano2008rich,zlatic2009rich}. Consider $\strength_{min}$ is the minimal weight linking two nodes in a network and the link between nodes $i$ and $j$ has a weight of $\strength_{ij}$. This link is represented by $\lceil \strength_{ij}/\strength_{min} \rceil$ links and the ranking is performed in units of the minimal weight. Each node is assigned to $\nodeWeight_i=\lceil\sum_j\strength_{ij}/\strength_{min}\rceil$ links, and part of this quantity arisen from the node's linkage to nodes of a higher rank is referred to as $\nodeWeightPlus_r$; similarly, the remaining proportion, $\nodeWeight_r-\nodeWeightPlus_r$, is the normalized weight that node $r$ shares with nodes of a lower rank. The core boundary is node $r^*$ such that $\nodeWeightPlus_{r^*} > \nodeWeightPlus_{r^*+i}$ for $i=1,\ldots N-r$.

Similarly, the notion of a rich-core can equally be applied to directed networks by dividing links into in-- and out-- links and quantifying their corresponding weights. In this case the definition of a rich-core does not only depend on the weight of the nodes but also the direction of their links (in-- or out--degrees). An example is the assessment of web pages using PageRank where it has been observed that the popularity of a node is closely related to its in--degree~\cite{Fortunato2}. Let 
$\nodeWeightPlusIn_r + \nodeWeightPlusOut_r$
be the total strength of interactions between node $r$ and the nodes $r'<r$, the core boundary is the node $r^*$ such that 
$\nodeWeightPlusIn_{r^*}+\nodeWeightPlusOut_{r^*} > \nodeWeightPlusIn_{r^*+i}+\nodeWeightPlusOut_{r^*+i}$ 
for $i=1,\ldots N-r$.

\begin{figure}[tb]
\begin{center}
\includegraphics{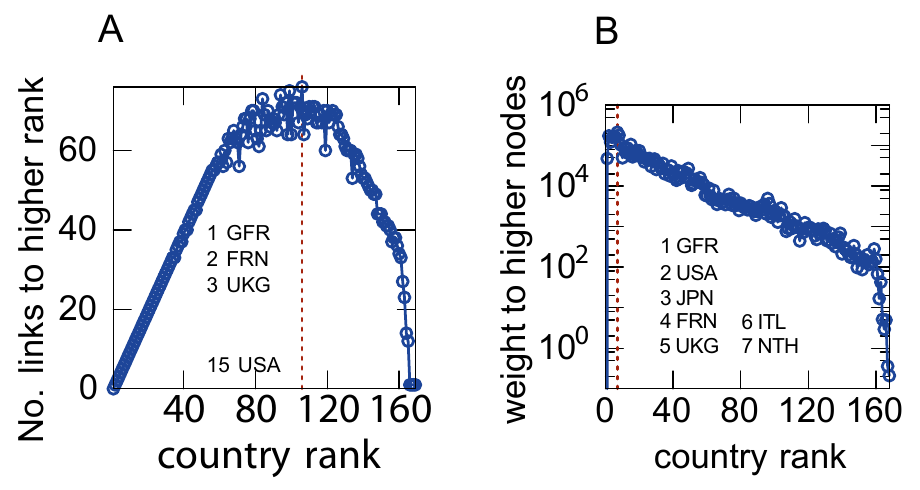}
\end{center}
\caption{\label{fig:weightedDirectedCore} The World Trade network in year 1990. (A) The unweighted undirected network which represents trade relationships between the countries. The core has 106 countries. (B) The weighted directed network representing the exchange of wealth between the countries. The core has seven countries.}
\end{figure}

We analyzed the World Trade network \cite{gleditsch2002expanded} in which nodes are countries and links are trade channels; the latter can represent the direction of trade to specify an import or export relationship. The associated financial value can be seen as the weight of a given link. The overall connectivity of the network is found to be high as countries are interrelated in many ways. We first examined the network as a binary and unweighted network and a link simply refers to the presence of trade. A total of 60\% of all the countries in the world are part of this network as a result of the globalization of trading at the time \cite{worldtrade2013}. By ranking countries in descending order of their degree, Fig.~\ref{fig:weightedDirectedCore}\emph{A} shows the number of trade relationships each of these countries has with countries of a higher rank in 1990 and the core consists of 106 countries. This interpretation of the network confirms close trading relationships found among the countries within the European Union (EU), which are highly ranked. The USA, which is a dominant importer and exporter by value, is ranked 15. Now, if we take both the direction and weight into consideration and nodes are ranked in the descending order of their exports (Fig.~\ref{fig:weightedDirectedCore}\emph{B}), members found in the core are consistent with the top importers and exporters in the world at the time \cite{worldtrade2013}.

\subsection{Evolution}
\begin{figure}[t]
\begin{center}
\includegraphics{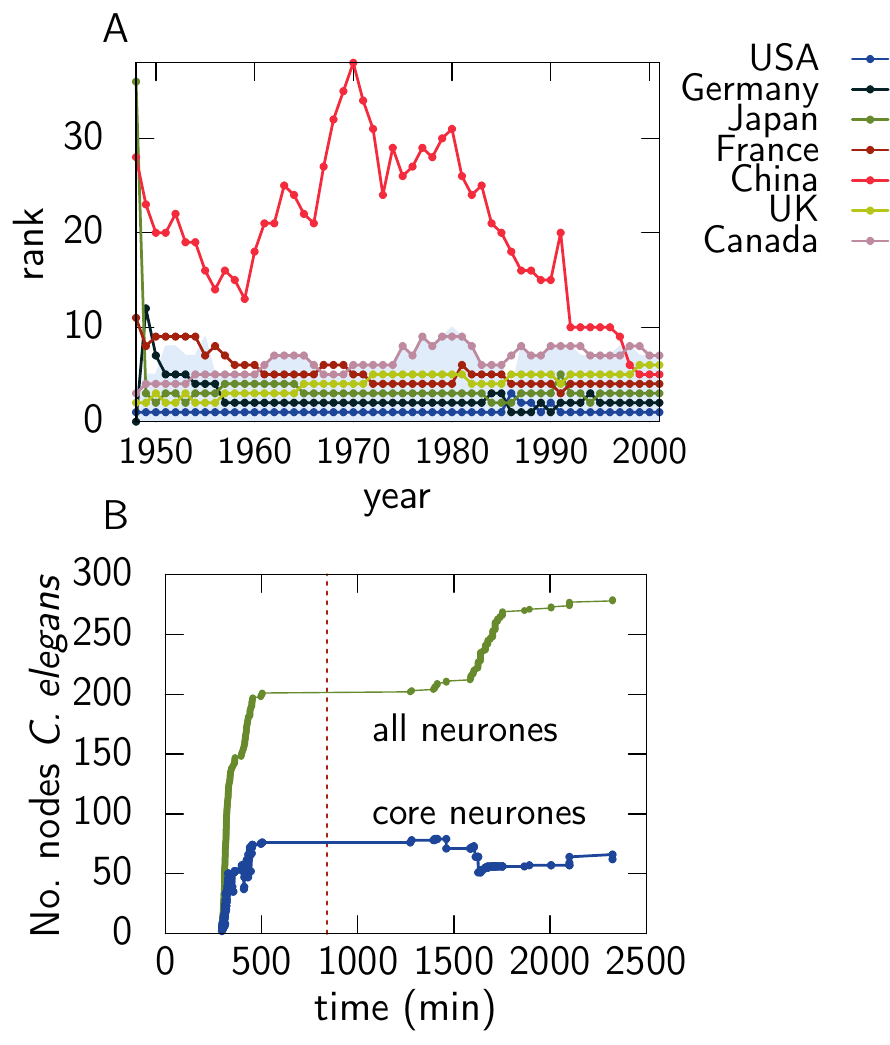}
\end{center}
\caption{\label{fig:coreEvolve} (A) Evolution of the core for the World Trade network. The shaded area (light blue) is the size of the core. The countries shown are the members of the core in year 2000.  (B) Evolution of the core for \emph{C.~elegans}. The red dotted line marks the hatching time.  }
\end{figure}

Networks are often found to be temporal in nature as they are subject to formation, dissolution and rewiring of links \cite{leskovec2008dynamics, kostakos2009temporal}. Continuing with the example of World Trade we studied the core in the period between 1948 to 2000. Applying our method results in a core which consists of very selective group of countries, corresponds to 4\% to 6\% of all the countries (Fig.~\ref{fig:coreEvolve}\emph{A}). This can be explained by referring to the way in which the World Trade has grown since the Second World War. International trading is said to be growing steadily but unevenly since the 1940s, as trade barriers were imposed by events such as the Cold War. The network is also strongly influenced by other key historical events, geographical distance, composition (e.g. products and services) and nature of trade \cite{worldtrade2013}. Throughout the 1980s and 1990s, we see substantial reduction in the cost of shipping due to the explosion of air freight, the collapse of the Soviet Union has led to many independent countries, and industrialization of developing countries; all these events have shaped the development of trade worldwide, leading to a great leap in globalization. This is somewhat reflected by the way the membership of the core changes over time. The USA, Germany, Japan, France and the United Kingdom have been the top importers and exporters in the world during the period of the study and it can be seen that these countries have been members of the core for the entire time. Canada has been drifting in and out of the top ten in the World Trade ranking during the same period, and we can see a similar variability in its core membership. In addition, the economic reform in China, which started in the late 1970's, has led to a steady growth of {$\sim$}9\% in World Trade per year, and our results illustrate that China became a member of the core in 1997. In fact, China joined the World Trade Organization in 2001. 

Another example of how changes in a core tie in with key events in real networks can be found in the development of \emph{C. elegans} neuronal connections (Fig.~\ref{fig:coreEvolve}\emph{B}). Almost the entire core is developed within the first 500 minutes~\cite{towlson2013rich}. The formation of the core coincides with Embryogenesis, and it has been suggested that the highly connected neurons appear in the early development to minimize the energy cost by creating the core connections among key nodes that are not physically far part; these connections can then be extended during the process of body elongation. New neurons are found after hatching in the late L1 larval stage at approximately 1250 minutes and the total number of neurons continues to grow until the start of L4 larval stage at approximately 2400 minutes \cite{byerly1976life}. The post-hatching development causes the relative connectivity among the existing core neurons to decrease, resulting in a reduction in the overall size of the core. The shrinking of the core coincides with the timing of Gonadogenesis.

\subsection{Multi--core and Relative Size}
\begin{figure}[t]
\begin{center}
\includegraphics[width=6cm]{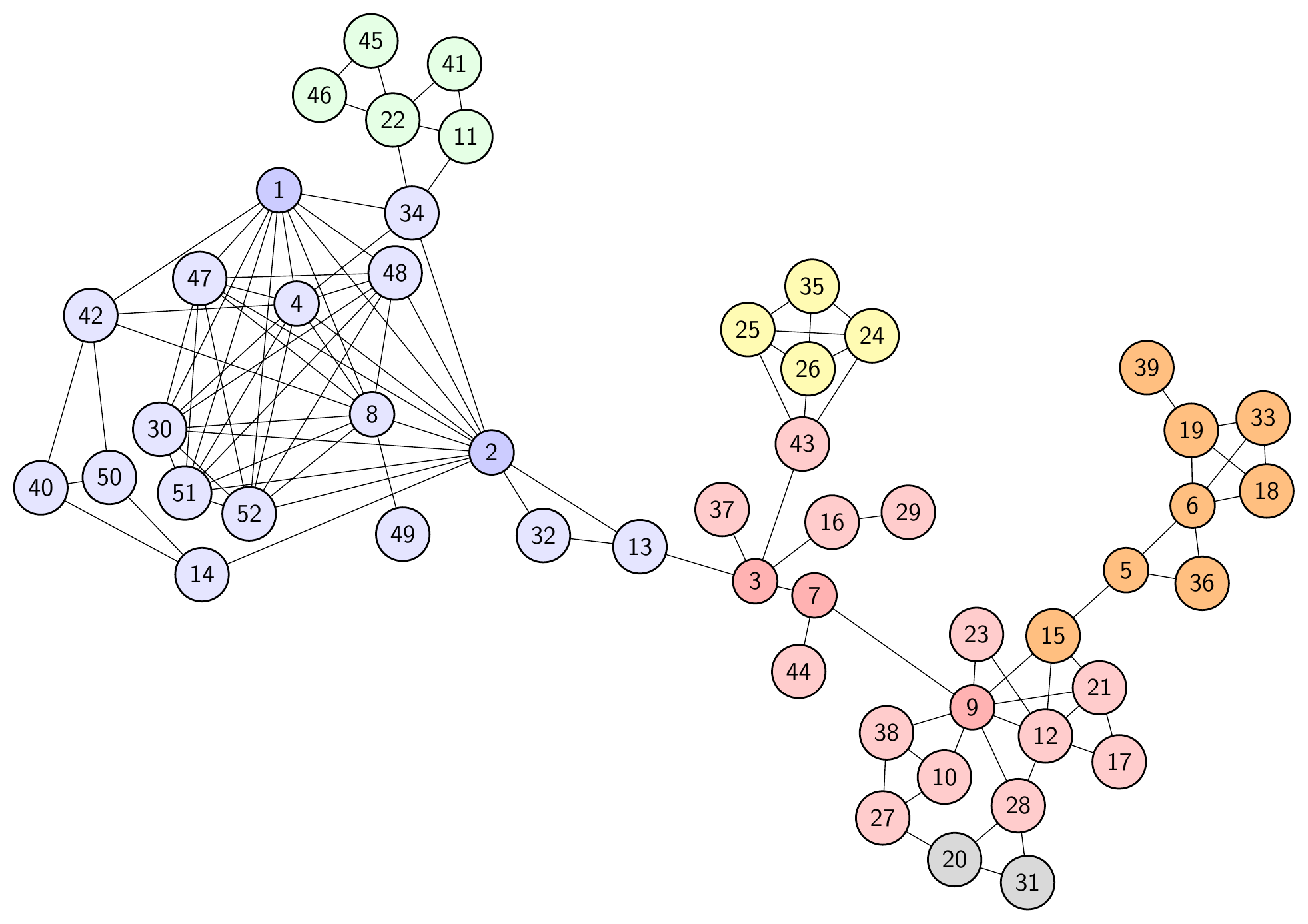}
\end{center}
\caption{\label{fig:multiCore}  Core of the NetSci network and its internal structure. The multi--core found is further subdivided into their corresponding parts and highlighted in different colors.} 

\end{figure}

A network can have a multi--core which consists of subgroups of nodes or disjoint parts. The former refers to a core which has a well-defined internal structure such that cliques can be observed within its configuration. A simple way to divide them into their prospective groups is to re--use the ranking of the nodes (Materials and Methods). We analyzed the network of scientists \cite{newman2006} (NetSci) by firstly defining the core of the network (Fig.~\ref{fig:multiCore}). Secondly, we examined the internal structure of the core by employing the algorithm for subgroup classification, and our results reveal that the core can be divided into well defined subgroups. On the other hand, the latter can be assessed using reachability and can be achieved using a shortest-path algorithm.



The size of a core is an important property of a network as it has been suggested that a sizable core makes a network more flexible and adaptable to changes~\cite{csete2004bow}, and a small core makes a network more controllable~\cite{liu2011}. The relative size of a core $c=N_c/N$ is the ratio between the number of nodes in the core $N_c$ relative to the total number of nodes in the network $N$. A network that has no periphery will have a relative core of $1$, e.g. a fully connected network; a star network with $N$ nodes has a relative core of $1/N$ and a core-less network has no links. 
%
%
We examined the relative size of a range of complex networks, ranging from 
man-made, social and biological, and the cores observed in these networks have different size and structure, in part reflecting their functionality. For instance, the Amazon and the Internet networks both have a relatively small core. The former is found to be disjoint, and as the network contains information about products recommendations, the results provide evidence of efficient information transfer within the network but only restricted to localized parts. The latter has a well connected single core which reflects its design for efficient routing between Autonomous System domains. The \emph{C.~elegans} neuronal network \cite{watts1998collective} has a relative large single core, and this perhaps reflects the adaptability of the neuronal network to living. The CA-road network \cite{leskovec2009community} has a relatively large multi--core which represents the existence of many crossroads, providing great flexibility in route choices as they present many possibilities between different geographical points. We did not observe any characteristic size of the core related to the kind of networks, i.e. man-made, social or biological. 

\begin{figure}[tb]
\begin{center}
\includegraphics{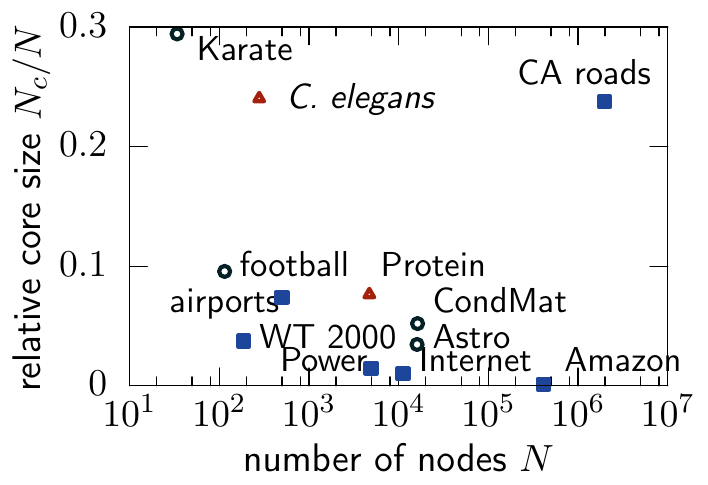}
\end{center}
\caption{\label{fig:Coreness} Relative size of the core in different kinds of real networks. Man-made (square): US airports (airports) \cite{colizza2007reaction}, Amazon recommendation (Amazon) \cite{LeskoAmazon}, Californian (CA) roads \cite{leskovec2009community}, Internet \cite{chen2002origin}, Power grid (Power) \cite{watts1998collective} and World Trade (WT) 2000 \cite{gleditsch2002expanded}. Social (circle): Astrophysics collaborations (Astro) \cite{newman2001structure}, Condensed Matter collaborations (CondMat) \cite{newman2001structure}, American College football (football) \cite{girvan2002community} and the Zachary Karate club (karate) \cite{zachary1977information}. Biological (triangle): \emph{C. elegans} \cite{watts1998collective} and Protein \cite{colizza2005characterization}.}
\end{figure}

\subsection{Anomalies in the core and its nodes} 
\begin{figure}[tb]
\begin{center}
\includegraphics{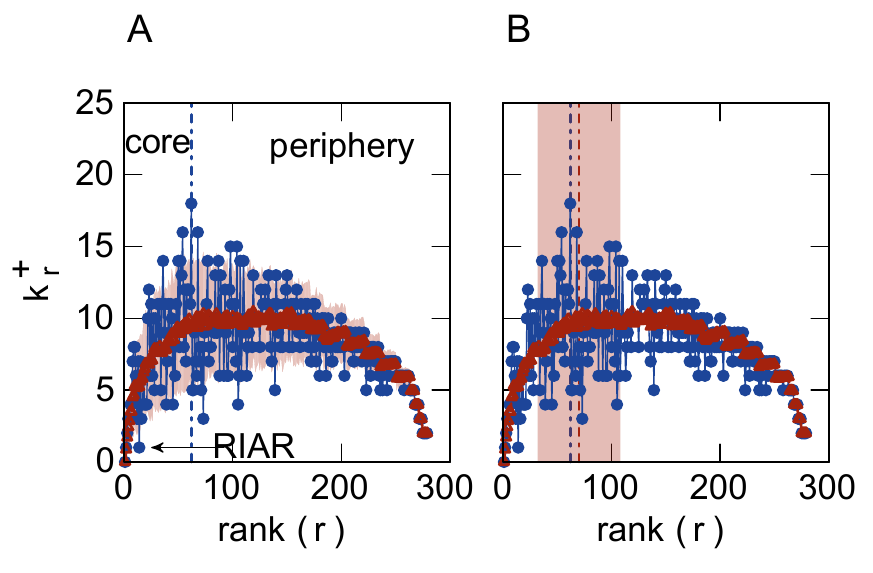}
\end{center}
\caption{\label{fig:compRC1} (A) Comparison on the connectivity of the core between the empirical data and the null model for \emph{C. elegans}. The shaded area marks two standard deviations from the sample mean. The neuron RIAR is found to have an anomalous connectivity with the highest ranked nodes. (B) Comparison  on the core size between the empirical data (blue line) and the null model (red line). Similarly, the shaded area shows two standard deviations.}
\end{figure}

Our rich-core method can be used to detect anomalies with respect to the core size and connectivity among members. We employed a randomization method to create an ensemble of networks which in turn are used as a reference null model for comparison purposes. The randomization is restricted to preserve the ranking of the nodes, that is the weight/degree of the nodes (Materials and Methods). 
Using \emph{C. elegans} as an example, we created an ensemble of 100 networks with the same degree distribution as the original network. From the ensemble we evaluated the average number of links that node $r$ has with nodes of a higher rank, i.e. $\langle \linksRC_r \rangle$, and the standard deviation of this quantity. Fig.~\ref{fig:compRC1}\emph{A}  shows the number of links between a node of rank $r$ and a node of rank $r'<r$ and the average number of links obtained from the null--model. The boundary of the core--periphery is the dotted blue line. The red line is $\langle \linksRC_r \rangle$ and the pink shaded area demarcates two standard deviations from the mean value. This implies that nodes outside the shaded area can be considered anomalous. For example, node 14 which corresponds to neuron RIAR, has an anomalous connectivity as it only shares one of its links with nodes of a higher rank.
The randomization can also be use to decide if the size of a core is within the expected value. Fig.~\ref{fig:compRC1}\emph{B} shows the average size of the core obtained from the ensemble of networks (red dotted line) and the pink area corresponds to two standard deviations from this mean. The dotted blue line is the size of the core obtained from the empirical data; as this value falls inside the pink area we can conclude that the size of the core for the \emph{C. elegans} is what we expected.

\section{Conclusions}

We developed a method to profile core/periphery in networks by theoretically coupling the notion of a rich-club and the persistence probability of random walks, resulting in finding a rich-core. The method is simple, fast and applicable to large networks, as demonstrated by the range of artificial, social and biological networks. Our theoretical work shows that the method is robust and can be extended to both directed and weighted graphs. Furthermore, we uncovered important characteristics in real networks through the profiling of the core/periphery structure, contributing towards the understanding of the significance of such meso-scale property. In particular, development and realignment of relationships are often found in networks with temporal nature, and our studies on both the World Trade and \emph{C. elegans} networks showed that the evolution of the core closely tie in with the timing of major historic events and key development stages respectively. We also demonstrated how the method reveals the organization of the core in a microscopic scale and characterizes multi-core. We showed an effective way to identify any anomalies in the core by comparing the membership of the core of a given network with its null counterparts, and how this approach can be used to evaluate the expected size of of a given core.

We thank J. Schormans for his comments on the manuscript, X. Lu and C. Gray for early discussions related to this research. 


\newpage

\section*{Methods}
\subsection*{Rich--club Coefficient and Core/Periphery Profile} 
{To find the boundary of a rich-core, we examine the escaping time it takes a random walker to leave a core. The escaping time is related to the notion of persistence probability $\profile$ which indicates the cohesiveness in a subgraph~\cite{della2013profiling, piccardi2011finding}. The persistence probability of cluster $S_C$ is
\begin{equation}
\profile_c = \frac{\sum_{i,j\in S_c} \pi_i m_{ij}}{\sum_{i\in S_c}\pi_i}.
\end{equation}
where $\pi_i$ is the probability that a random walker is in node $i$, and $m_{ij}$ is the probability that a random walker moves from node $i$ to node $j$.  
The escaping time is $\tau_c = (1-\profile_c)^{-1}$.
}
Assuming that $\profile_x$ can be approximated by a continuous function $\alpha(x) = g(x)/f(x)$ where $x$ is a continuous rank. 
The first derivative $\alpha'(x)$ increases with $x$, as the number of nodes rises and eventually diverges to 1 when all the nodes are included. The rate of increase in the persistence probability is given by the second derivative 
$\alpha''(x) = g''(x)/f(x) +2g(x)f'(x)^2/f(x)^3-2f'(x)g'(x)/f(x)^2-g(x)f''(x)/f(x)^2$. To first approximation  $\alpha''(x) \simeq 0$ if $g''(x)=0$ as $f(x)$ is a positive increasing function of $x$. 

\noindent The rich-club coefficient is defined as ~\cite{Zhou04}:
\begin{equation}
\phi(r) = \frac{2E(r)}{r(r-1)}= \frac{2\sum_{i=1}^{c} \linksRC_i}{r(r-1)}
\label{equ:richclub}
\end{equation}
where $E(r)$ is the number of links between the $r$ nodes.
For undirected networks $\profile_c$ is given by the sum of the links between the nodes in $S_c$ divided by the sum of the degrees of the nodes in $S_c$. If $a_{ij}$ are the elements of the adjacency matrix, then
\begin{equation}
\profile_c = \left( \sum_{i,j \in S_c} a_{i,j}\right)/\sum_{i\in S_c}\degree_i=\frac{\sum_{i=1}^{c} \linksRC_i}{\sum_{i=1}^{c} \degree_i}.
\label{equ:coreandp}
\end{equation}
Again, assuming that $\profile_c$ can be approximated with a continuous function $\profile_c=g(c)/f(c)$ where $g(x) = \int \linksRC(x) dx$ then $g''(x^*)=0$ means that $\linksRC(x^*)$ has a maximum or a minimum at the value $x^*$,  in this case we are interested in the maximum. We refer the point $x^*$ where $g''(x^*)=0$ as the boundary of the rich-core and nodes in $S_x$ are the members of the core. The explicit relationship between the escaping time and the rich-club coefficient is obtained by substituting Eq.~\textbf{\ref{equ:richclub}} into Eq.~\textbf{\ref{equ:coreandp}}, giving 
\begin{equation}
\profile_r = \frac{r(r-1)}{2} \frac{\phi(r)}{\sum_{i=1}^{r} \degree_i}.
\end{equation}

\noindent For weighted networks, as we are considering the weights as undirected multilinks the same argument applies when defining the core. For directed networks, the persistence probability $\profile_c$ is given by the ratio between the number of times a random walker transits inside the core and the number of times it visits the core. The former is proportional to the number of links inside the core, which we denoted as 
$\nodeWeightPlusIn_{r^*}+\nodeWeightPlusOut_{r^*}$.
 The latter is given by the sum $\sum_{i=S_c}\pi_i$. The in--degree $k_i$ of node $i$ is assumed to be a good approximation of $\pi_i$ ~\cite{Fortunato2}; hence, the number of times a random walker visits the core increases with $r$, as the nodes are ranked in decreasing order of their in--degree.


\subsection*{Rich-Core Algorithm} 
To find the core of a given network, 
\begin{enumerate}
\item rank nodes in decreasing order of their weight (specifically, degree, in--degree and weight for undirected, directed weighted networks respectively.) 
\item evaluate the number of links $\linksRC$ between node with rank $r$ and nodes with rank $r'<r$
\item find the boundary of the core, defined by the node where $\linksRC_{r^*} > \linksRC_r$ for $r>r^*$
\end{enumerate}

\subsection*{Rank degeneracy}
As numerous nodes can have the same degree, the ranking of  nodes with equal degree is not well defined. This degeneracy in the ranking scheme would affect the determination of the boundary nodes, and hence the size of a core. To evaluate the effect of the degeneracy in the definition of a core we randomly re-ranked the nodes with equal degree and measured the change of the core nodes. We observed that this re--ranking only has a minor effect when defining a core.




\subsection*{Multi--core}
In undirected networks the ranking of the nodes inside the rich-core can be used to divide the core nodes into subgroups. The algorithm is based on the idea that a random walker that is in a  core node of low rank will ``drift"  to a node of a higher rank, the algorithm evaluates this process in reverse. 
\begin{enumerate}
\item Starting from the highest ranking node
\item If the node is not a member of a subgroup, create a new subgroup
\item All the neighbors of the selected node which have a lower rank than the selected node belong to the same subgroup. 
\item Repeat this until there is no neighbors with a lower rank left. Mark these nodes as members of the same subgroup. 
\item If all the nodes of the rich-core have been marked then finish, if not
\item Select the next ranking node, go to step 2
\end{enumerate}
Inside a core there can be nodes which are at the boundary of two subgroups. To decide which subgroup these nodes belong to, we assign the boundary nodes to the subgroup in which they share the highest number of links. If the nodes share the same number of links with both subgroups, the subgroup with the highest ranked nodes will be chosen. 


\subsection*{Construction of Null Models}
The null model is generated using Zlatic~et. al approach \cite{zlatic2009rich} which is a generalization of Maslov, Sneppen and Zaliznyak method (MSZ)~\cite{Maslov2004} to generate null models. Zlatic~et~al. redistributes the weights of the links by preserving the strength of the nodes as follows. If $\weight_m$ is the minimum value of the weights in the original network then the rewiring is done by changing the weights of two pairs of links by an amount $\weight_m$. The rewiring consists of selecting two links at random and exchanging one of the end nodes of the first link with an end node of the second link.


%

\end{document}